\documentclass{aa}
\usepackage{txfonts}
\usepackage{graphicx} 

\newcommand{\diff}[2]{\frac{\partial #1}{\partial #2}}
\newcommand{\difff}[2]{\frac{\partial^2 #1}{\partial #2^2}}

\newcommand*{\eps}{\epsilon}

\newcommand{\ele}{E}
\newcommand{\phe}{\varepsilon}

\begin{document}
%

\title{Electron Acceleration in Solar Flares: \\
Theory of Spectral Evolution}

\titlerunning{Theory of Spectral Evolution in Solar Flares}

\author{P. C. Grigis \and A. O. Benz}

\offprints{\\Paolo C. Grigis,
\email{pgrigis@astro.phys.ethz.ch}}

\institute{Institute of Astronomy, ETH Zurich, 8092 Zurich, Switzerland}

\date{Received  xxx / Accepted yyy}

\abstract
{Stochastic acceleration is thought to be a key mechanism in the energization of
  solar flare electrons.}
{We study whether stochastic acceleration can reproduce the observed
  soft-hard-soft evolution of the spectral features of the hard X-ray emitted by
  suprathermal electron. We pay special attention to the effects of particle
  trapping and escape.}
{The Fokker-Planck equation for the electron distribution is integrated
  numerically using the coefficients derived by Miller et al. for transit-time
  damping acceleration. The electron spectra are then converted to photon
  spectra for comparison with RHESSI observation of looptop sources.}
{The presence of particle escape softens the model spectra computed in the
  stochastic acceleration framework. The ratio between the efficiency of
  trapping and acceleration controls the spectral evolution which follows a
  soft-hard-soft pattern. Furthermore, a \emph{pivot point} (that is, a common
  crossing point of the accelerated particle spectra at different times) is
  found at around 10~keV. It can be brought into agreement with the observed
  value of 20~keV by enhanced trapping through an electric potential.}
{The model proposed here accounts for the key features observed in the spectral
  evolution of hard X-ray emission from looptop sources.}


\maketitle

%
\section{Introduction}
%

Solar flares emit large amounts of continuum hard X-ray bremsstrahlung from
energetic electrons up to several hundreds of keV. The thermal energy of the
ambient electrons in the corona is substantially lower, with electron
temperatures observed up to 6~MK in nonflaring active regions (Brosius et al.
\cite{brosius96}, Benz \& Grigis \cite{benz02}). Clearly, a mechanism is needed
to accelerate the electrons out of the thermal distribution up to relativistic
energies.  More than $10^{31}$~erg (Lin et al. \cite{lin03}, Emslie et al.
\cite{emslie04}, Kane et al.  \cite{kane05}) of energy can be transferred into
non-thermal electrons during a major solar flare over a period of a few hundred
seconds.

These numbers set stringent requirements to acceleration models, and it is by no
means a simple task to identify the physical processes involved. Most of the
proposed models fall into three broad classes: electric DC field acceleration,
stochastic acceleration and shock acceleration (see e.g. the review by
Aschwanden \cite{aschwanden02}).

Miller et al. (\cite{miller96}, hereafter MLM) proposed a stochastic
acceleration mechanism where electrons are energized by small amplitude
turbulent fast-mode waves, the transit-time damping model. MLM showed that their
model could successfully account for the observed number and energy of electrons
accelerated above 20 keV in subsecond spikes or energy release fragments in
impulsive solar flares. However, they made no attempt to explain the observed
hard X-ray spectra (which are softer than predicted by the transit-time damping
model) and did not consider spectral evolution. The MLM approach does not
account for escape and transport processes.

What is the effect of escape on the electron spectrum in the accelerator? In
stochastic acceleration, each electron describes a random walk in energy space.
The effect of escape is to shorten the dwell time of the particles in the
accelerator, and therefore it limits the average energy of the electrons. This
results in a softer spectrum. Therefore, we need to take in account escape to
compare quantitatively the model predictions with hard X-ray observations.
Furthermore, transport effects account for the modification of the electron
spectrum from the time they leave the accelerator until they reach the hard
X-ray emitting regions. The usual interpretation is that the particles are
accelerated near the top of magnetic loops, but most of the X-rays are emitted
in the loop footpoints, where the density is much larger.  Nevertheless, looptop
sources are also observed (Masuda et al.  \cite{Masuda94}).

Observations from the Reuven Ramaty High Energy Imaging Spectrometer (RHESSI,
Lin et al. \cite{lin02}) deliver hard X-ray spectra of both footpoint and
looptop sources (Emslie et al. \cite{emslie03}, Liu et al. \cite{liu04}).
Looptop sources are better suited for comparison with accelerator models than
footpoint sources, since one does not need to take into account the spatial
transport from the accelerator. The emission from the modeled electron
distribution can be directly compared with the observed spectra of the looptop
source.

%
%
\begin{table}
\caption{Observational constraints for looptop sources
  (taken from Battaglia \& Benz \cite{battaglia06}).}
\label{tab:obsres}
\centering
\begin{tabular}{l l l}
\hline\hline
Parameter Description & Average & Range\\
\hline
Pivot-point energy $\phe_*$ & 20 keV & 16--24 keV\\ 
Pivot-point flux$^\mathrm{a}$ $J_*$ &  2 & 1--4  \\
Mean temperature & 22 & 18--25 MK\\
Nonthermal fitting range: \\
\hskip 0.5cm Lower energy $\phe_\mathrm{min}$ & 25 keV & 20--30 keV\\
\hskip 0.5cm Upper energy $\phe_\mathrm{max}$ & 60 keV & 40--80 keV\\
\hline 
\end{tabular}
\begin{list}{}{}
\item[$^{\mathrm{a}}$]In units of  photons cm$^{-2}$ s$^{-1}$ keV$^{-1}$.
\end{list}
\end{table}
%
%

Battaglia \& Benz (\cite{battaglia06}) present a careful and comprehensive study
of the spectral evolution of looptop and footpoint sources for 5 well observed
RHESSI flares.  They find that looptop sources systematically show the
soft-hard-soft (SHS) behavior observed spectroscopically for the total flare
emission (Grigis \& Benz \cite{grigis04} and references therein). On the other
hand, some footpoint sources do not follow the SHS pattern. These observations
suggest that the SHS behaviour is a property of the acceleration process.

The observed SHS behavior manifests itself as a tight correlation between the
curves describing the time evolution of the negative spectral index $\gamma$ and
the normalization $J_{E_0}$ of the non-thermal power-law component of the photon
spectra, measured at a fixed reference energy $E_0$. RHESSI observations show a
linear relation between $\gamma$ and $\log J_{E_0}$ (Grigis \& Benz
\cite{grigis04}). This is equivalent (Grigis \& Benz \cite{grigis05}) to the
presence of a \emph{pivot point}, that is, a common point where the power-law
component of the photon spectra observed at different times intersect.
Therefore, the position of the pivot point can be found observationally by
fitting the regression line to $\gamma$ and $\log J_{E_0}$. Since the
observations show that during the impulsive phase of most flares the scatter
around such a model is relatively small, such fittings provide us with two new
meaningful observational parameters which can be used for comparison with
theories. The best-fit pivot-point energy can be measured by RHESSI with an
accuracy of 10--20\% (Battaglia \& Benz \cite{battaglia06}).

The measured energies of the pivot points lie in the range of 18--24 keV for
looptop sources and 13--15 keV for footpoint sources. In Table \ref{tab:obsres}
we summarize the observational results from Battaglia \& Benz
(\cite{battaglia06}) which we will use to constrain the transit-time damping
acceleration model.

In this paper we study how well the transit-time damping model endowed with an
escape mechanism can account for the hard X-ray spectra of the looptop sources
seen by RHESSI and, in particular, for their spectral evolution. We investigate
whether the spectrum of the photons emitted by the modeled electrons in the
accelerator shows SHS behavior and how well this quantitatively agrees with the
observations.

We proceed as follows: a modified version of the MLM model with a better
characterization of electron trapping is presented in Sect. \ref{model}.  The
evolution of the electron energy distribution function is given by a diffusion
equation which is integrated numerically and transformed into photon spectra. In
Sect. \ref{results} we present the resulting photon spectra and compare their
behaviour with the observations. The results are discussed in Sect.
\ref{discussion} and conclusions are drawn in Sect. \ref{conclusion}.

%
\section{Pivot-point theory}
\label{pivpointintro}

%
%
\begin{figure*}
\resizebox{\hsize}{!}{\includegraphics{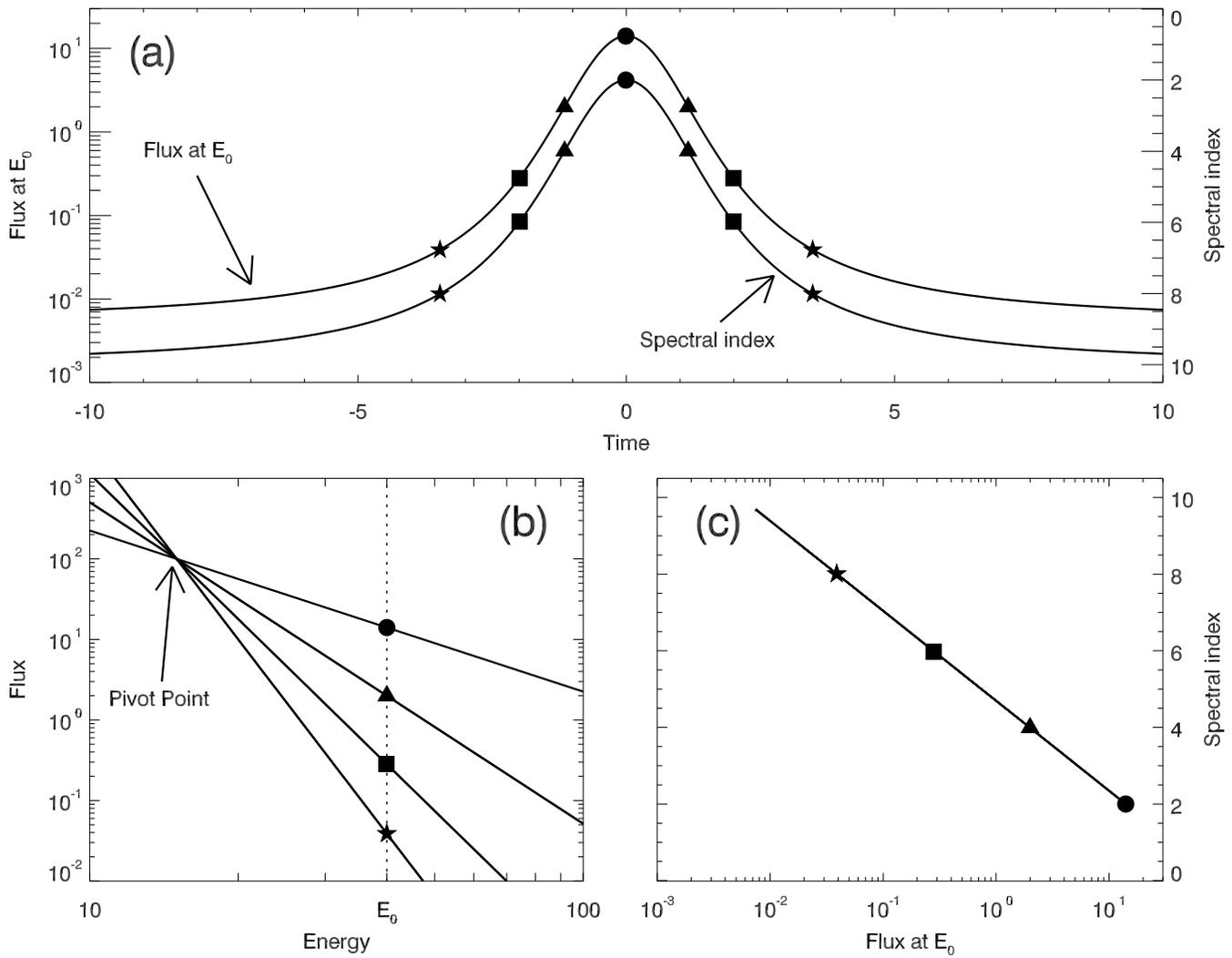}}
\caption{
Schematic illustration of the basic properties of a spectral evolution.
\emph{Panel (a)}: time evolution of the logarithm of the flux measured
at reference energy $E_0$ (upper curve) and of the spectral index (lower curve)
in the case where they are perfectly correlated.
\emph{Panel (b)}: Spectra for different values of the spectral index, crossing
at the pivot point.
 \emph{Panel (c)}: Linar dependence of the spectral index on the logarithm of
 the flux. In all panels, the stars, squares, triangles and circles mark points with
spectral indices of, respectively, 8, 6, 4 and 2. 
}
\label{fig:pivpoint}
\end{figure*}
%
%

  In this section the mathemathical foundations needed for the comparison of the
  model results with the observations are presented. In particular we explain
  what are the consequences of the correlation in time between flux and spectral
  index. By the introduction of a new parameter, the pivot-point, the time
  dependence can be eliminated, thus simplifying the analysis.

  The theory is presented in a general context, and can be applied to any kind
  of spectra (for example, photons or electrons) described by a power-law in
  some energy range. The starting point of this analysis is the presence of a
  correlation between the logarithm of the flux, $\log F_{E_0}$, measured at
  energy $E_0$ and the spectral index $\delta$, where the time dependent
  spectrum is given by
\begin{equation}
\label{tdepsp}
  F(E,t)=F_{E_0}(t)\left(\frac{E}{E_0}\right)^{-\delta(t)}\,,
\end{equation}
for $t$ in a time interval $T_\mathrm{INT}$ (for instance, the duration of a
flare emission peak).  Panel (a) of Fig.  \ref{fig:pivpoint} shows the ideal
case where the correlation in time between $\delta$ (lower curve) and $\log
F_{E_0}$ (upper curve) is perfect: the two curves are parallel. In such a case,
a plot of $\delta$ vs. $\log F_{E_0}$ is linear. This shown in panel (c). The
relation between these two parameters can be expressed as
\begin{equation}
\label{pivdep}
\delta(t)=a\cdot\log F_{E_0}(t)+b\,,
\end{equation}
where $a\not=0$, and the parameters $a$ and $b$ do not depend on time. 

Some points with selected values of $\delta$ are marked by special symbols in
Fig. \ref{fig:pivpoint}: circles for $\delta=2$, triangles for $\delta=4$,
squares for $\delta=6$ and stars for $\delta=8$. In panel (b) the spectra
corresponding to these values of the spectral index are plotted together. It is
evident that all these spectra intersect in a common point, the \emph{pivot
  point}. The coordinates of the pivot point are defined as $(E_*,F_*)$. The
presence of a linear relation between $\delta(t)$ and $\log F_{E_0}(t)$ is
equivalent to the presence of a pivot-point. This means that panel (a) in
Fig. \ref{fig:pivpoint} implies both (b) and (c). A detailed proof follows.

\subsection{Proof of the equivalence between pivot point and correlation of
  $\delta$ with $\log F$}

The following statements are equivalent:
\begin{itemize}
\item[(i)] There is an $E_*\not=E_0$ such that $F(E_*,t_1)=F(E_*,t_2)=:F_*$ for
  all $t_1,t_2 \in T_\mathrm{INT}$.
\item[(ii)] There exist constants $a\not=0$ and $b$ such that
\begin{equation}
\label{deltalindep}
\delta(t)=a\log F_{E_0}(t)+b
\end{equation}
for all $t\in T_\mathrm{INT}$.
\end{itemize}
The parameters $a$ and $b$ are given by
\begin{eqnarray}
\label{abdef}
a & = & \frac{1}{\log (E_*/E_0)}\,, \\
b & = & \frac{-\log{F_*}}{\log (E_*/E_0)}\,.
\end{eqnarray}
We give first the proof that (i) $\Longrightarrow$ (ii). Because $F(E_*,t)=F_*$
for all $t\in T_\mathrm{INT}$, Eq.~(\ref{tdepsp}) yields
\begin{equation}
\log F_*=\log \left[F_{E_0}\left(\frac{E_*}{E_0}\right)^{-\delta}\right]=\log F_{E_0} -\delta\log{(E_*/E_0)}\,
\end{equation}
which can be solved for $\delta$, yielding
\begin{eqnarray}
\delta & = & \frac{\log F_{E_0}}{ \log{(E_*/E_0)}}-\frac{\log F_*}{\log{(E_*/E_0)}}\\
       & = & a\log F_{E_0}+b\,,
\end{eqnarray}
where $a\not= 0$ for finite values of $E_*$,$E_0$.\newline
Now we prove (ii) $\Longrightarrow$ (i). Let us define $E_*=E_0\exp (1/a)\not=E_0$.
Then using Eqs. (\ref{tdepsp}), (\ref{abdef}) and (\ref{deltalindep}) we get 
\begin{eqnarray}
  \log F(E_*) & = & \log F_{E_0}-\delta\cdot\log\left(\frac{E_*}{E_0}\right) \\
              & = & \log F_{E_0}-\frac{\delta}{a} \\
              & = & -\frac{b}{a}\,,
\end{eqnarray}
which does not depend on $\delta$ or $F_{E_0}$. Therefore, $(E_*,F_*)$ are the
coordinates of the pivot point. Q.E.D.

%
%
\begin{figure}
\resizebox{\hsize}{!}{\includegraphics{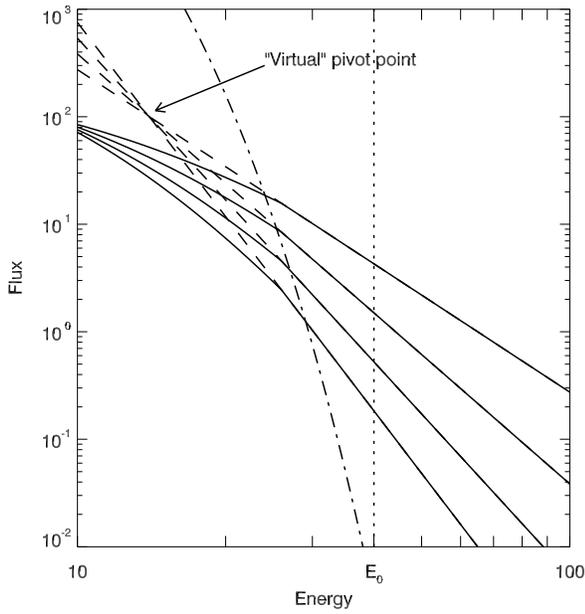}}
\caption{Spectra (mock data, continuous lines) turning over at low energy. The
  power-law extensions of the spectra (dashed lines) intersect in a
  \emph{virtual} pivot point, with larger flux than the one from the spectra.
  The dashed-dotted line represent an additional thermal component, effectively
  hiding the spectral behavior at low energies.
}
\label{fig:virtpivpoint}
\end{figure}
%
%

\subsection{Implications} 

  We have established above formally the equivalence between the presence of
  a pivot point and the linear relation between the spectral index and the
  logarithm of the flux measured at some fixed energy. We note that the shape of
  the curves describing the time evolution of $\delta$ and $F_{E_0}$ does not
  influence the results described above. The time dependence can be eliminated:
  this simplifies the analysis and allows the new parameter, the pivot-point
  position, to be used in the comparison of observations with theoretical
  models. Conversely, if the observation show that $\delta$ and $\log F_{E_0}$
  do not correlate well in time, we will not find any pivot-point like behavior
  in the spectral evolution.

  Our derivation assumes that the spectra are represented by power-laws at all
  energies. In this case, at energies lower than the pivot-point energy, the
  linear relation between flux and spectral index changes sign and higher fluxes
  would correspond to steeper spectra instead of harder ones. However, the
  observed spectra are not given by power-laws at low energies, because steep
  spectra are not integrable and therefore would imply an infinite number of
  particles, which is not physically possible. Therefore a deviation from the
  power-law form is expected in the form of a turnover at low energies
  (Saint-Hilaire \& Benz \cite{saint-hilaire05}, Sui et al. \cite{sui05}). This
  means that it is well possible that the spectra do not extend all the way to
  the pivot point energy, even if a correlation between the spectral index and
  the flux is observed at higher energies. In such a situation the pivot point
  is \emph{virtual}, in the sense that its flux is higher than the spectral flux
  at the same energy. In this case the analysis given above retain its validity,
  but the spectra used should be understood as the power-law extensions of the
  physical spectra. Such a situation is represented in Fig.
  \ref{fig:virtpivpoint}.

We emphasize that the presence of a pivot-point is a \emph{local} property of a
family of curves at energy $E_0$, in the sense that it can be seen as a
convergence point for the tangents in $E_0$ to the family of curves (in
logarithmic space). For power-law functions, the coordinates of the pivot point
are independent of the energy $E_0$, because their logarithmic derivative is
constant as a function of energy. Other functions exist for which there is a
different pivot point for each value of $E_0$.  

For the sake of completeness, we report here the differential equation whose
general solution is the family of curves yielding a pivot point (in the sense of
a convergence point for the tangents in linear space) in $x^*(x),y^*(x)$:
\begin{equation}
\label{diffeqpiv}
  \frac{dy}{dx}+\frac{y}{x^*(x)-x}=\frac{y^*(x)}{x^*(x)-x}\,,
\end{equation}
where in our case, $x=\log E$, $x^*=\log E^*$, $y=\log F$ and $y^*=\log F^*$.

In the special case $x^*=\mathrm{const.}$ and $y^*=\mathrm{const.}$ (that is, a
unique pivot point for all energies), the general solution of Eq.
(\ref{diffeqpiv}) is given by $y-y^*=m\cdot(x-x^*)$, the family of all straight
lines intersecting at $x^*$ and $y^*$.

%

%
\section{The Model}
\label{model}
%
The transit-time damping model is described in great detail by MLM. Here we
first summarize the important features of the model presented by MLM, and then
describe our modification of the model.

\subsection{The Transit-Time Damping Model}

The source of the flare energy is an event of magnetic reconnection, where
magnetic field energy is transferred, by an unspecified mechanism, into
small-amplitude magnetoacoustic fast mode waves. These waves have very large
wavelengths, comparable with the length scale of the global magnetic field
restructuring induced by the reconnection process.

A cascading process then takes place, producing waves with larger wave vector
$k$ and isotropic distribution. This process transfers wave energy density from
the low $k$ into the large $k$ waves. The electrons interact with the parallel
component of the magnetic field of the large $k$ waves. An electron of velocity
$v$ can efficiently exchange energy with the waves if the resonance condition
$k_\parallel v_\parallel=\omega=k v_\mathrm{A}$
is satisfied, where the subscript $\parallel$ denotes the component parallel to
the ambient magnetic field, $\omega$ is the wave frequency and $v_\mathrm{A}$ is
the Alfv\'en speed. The same condition can also be expressed as
$\mu\eta v=v_\mathrm{A}$,
where $\mu$ and $\eta$ are the cosines of the angle between the ambient magnetic
field and the direction of propagation of the electrons and the waves,
respectively. From this form of the equation it is clear that only electrons
faster then $v_\mathrm{A}$ can be accelerated by this mechanism. The
perpendicular velocity $v_\perp$ is not affected by these interactions.

Since the acceleration only occurs in the direction of the ambient magnetic
field, the electron distribution becomes strongly elongated in parallel
direction in the absence of pitch-angle scattering. This reduces considerably
the efficiency of transit-time damping acceleration, because for fast particle
with $v\gg v_\mathrm{A}$ and $|\mu|\approx 1$ the resonance condition can only
be satisfied by waves with $|\eta|<v_\mathrm{A}/v$, which represent only a small
fraction of all the available phase space. The acceleration efficiency can
therefore be increased by the presence of a mechanism which isotropizes the
electron distribution, allowing particles with small $|\mu|$ to interact with
waves with $|\eta|>v_\mathrm{A}/v$. MLM assume that an isotropization mechanism
is present, but do not address the details of the process.

The model described by MLM is self consistent in the sense that it describes the
evolution in time of both the electrons (which are accelerated) and the waves
(which are damped). The efficiency of the electron acceleration by the waves is
determined by the total energy density of the waves $U_\mathrm{T}$ and by the
shape of the wave spectrum: large $k$ waves exchange energy with the electrons
faster because the transit time between the magnetic perturbations gets shorter.
The coefficients describing the diffusion of the electrons in energy space depend
on the product of $U_\mathrm{T}$ and $\langle k\rangle$, where the latter
represents the average wave-vector of the waves.

The evolution of the electrons is described by the Fokker-Planck equation for
the electron distribution in energy space. MLM consider the effects of the
acceleration by the waves and of collisions with the ambient plasma.

\subsection{Transit-Time Damping and Escape}

In this paper, we study the electron and photon spectra produced by a simplified
transit-time damping acceleration model. We do not address here the problem of
the time it takes for the accelerated electron distribution to reach
equilibrium, considering that MLM have established this time to be sufficiently
short. Therefore we need not to account explicitely for the MHD cascading
process, but we will instead keep $U_\mathrm{T}\cdot\langle k\rangle$ as a free
parameter of our model. This means that only one partial differential equation
describes the time-evolution of the system instead of a set of two coupled
equations.

A short description of the physical parameters upon which the model depends is
given in Table \ref{tab:vardesc}, where a default reference value for each
parameter is listed. In the following these default values will be implied
if different values are not explicitely mentioned.

%
%
\begin{table}
\caption{Description of the parameters used in the model.}
\label{tab:vardesc}
\centering
\begin{tabular}{l l l}
\hline\hline
Param. & Description & Default value\\
\hline
$B$ & Ambient magnetic field strength &  500 G\\ 
$n$ & electron density & $10^{10}$ cm$^{-3}$\\ 
$T$ & Ambient plasma temperature & 10 MK\\ 
$\log\Lambda$ & Coulomb logarithm & 18 \\
$\Omega_\mathrm{H}$ & Proton gyrofrequency & 4.78 MHz\\
$T_\mathrm{H}$ & Time unit $\Omega_\mathrm{H}^{-1}$ & $2.09 \cdot 10^{-7}$ s \\ 
$v_\mathrm{A}$ & Alfv\'en speed & $0.034c$ \\
$U_\mathrm{T}$ & Energy density of the accelerating \\
& waves & 2 erg cm$^{-3}$\\ 
$U_\mathrm{B}$ & Energy density of the ambient \\
& field ($\frac{1}{8}B^2/\pi$) & $10^4$ erg cm$^{-3}$\\
$\langle k\rangle$ & Average wavenumber of the \\
& turbulence & $4.8\cdot 10^{-4}$ cm$^{-1}$\\
$I_\mathrm{ACC}$ & Acceleration parameter: & (free) \\
& $I_\mathrm{ACC}=\displaystyle\frac{U_\mathrm{T}}{U_\mathrm{B}}\cdot
\frac{c\langle k\rangle}{\Omega_\mathrm{H}}$ & \\
$\tau$ & Escape time & (free) \\
\hline 
\end{tabular}
\end{table}
%
%

The transit-time damping acceleration model describes the evolution of the
electron population in energy and time. We denote by $N(E)\,dE$ the number of
electrons per cubic centimeter with energy in the interval $dE$ around $E$,
where $E$ is the dimensionless particle energy in units of $m_ec^2$. The total
electron density is $n=\int_0^\infty N(E)\,dE$. The evolution in time of the
distribution function $N(E,t)$ is given by the convective-diffusive equation

\begin{eqnarray}
  \label{eq:maindiff}
\diff{N}{t} & = & \frac{1}{2}\difff{}{E}\Big[\left(D_\mathrm{COLL}+D_\mathrm{T}\right)N\Big]
            -\diff{}{E}\Big[ \left(A_\mathrm{COLL}+A_\mathrm{T}\right)N \Big]\\
\rule[-3mm]{0mm}{8mm} &  & -S(E)\cdot N + Q(E) 
\end{eqnarray}

The coefficients $D_\mathrm{T}$ and $A_\mathrm{T}$ describe the diffusion and
convection in energy space due to the interaction of the electrons with the
waves. The coefficients $D_\mathrm{COLL}$ and $A_\mathrm{COLL}$ represent the
effect that collisions against the ambient plasma have on the distribution
function. We include the effects of the escape by a sink term $S(E)$ and a
source term $Q(E)$ described below.

The acceleration coefficients are given by MLM, and we report them here written
in dimensionless form:
\begin{eqnarray}
  A_\mathrm{T}(E) & = &
  \frac{\pi}{4}\,\beta_\mathrm{A}^2\,I_\mathrm{ACC}\,\gamma\beta g(\beta)\\ 
  D_\mathrm{T}(E) & = & 
  \frac{\pi}{8}\,\beta_\mathrm{A}^2\,I_\mathrm{ACC}\,\gamma^2 \beta^3 f(\beta)
\end{eqnarray}
where
\begin{eqnarray}
  \gamma & = & 1+E\,, \qquad  \beta=\sqrt{1-\gamma^{-2}}\,,
\qquad \xi= \frac{\beta_\mathrm{A}}{\beta} \\ 
  f(\beta) & = & -1.25-(1+2 \xi^2)\log \xi+\xi^2+0.25\xi^4 \\ 
  g(\beta) & = & \frac{1}{4\gamma^2} (4\xi^2\log\xi-\xi^4+1)+f(\beta)
\end{eqnarray}
The acceleration coefficients vanish for $E<E_\mathrm{A}$, where $E_\mathrm{A}$
is the kinetic energy of a particle with speed
$\beta_\mathrm{A}=v_\mathrm{A}/c$, because the resonance condition cannot be
satisfied below the Alfv\'en speed.

\begin{figure}
\resizebox{\hsize}{!}{\includegraphics{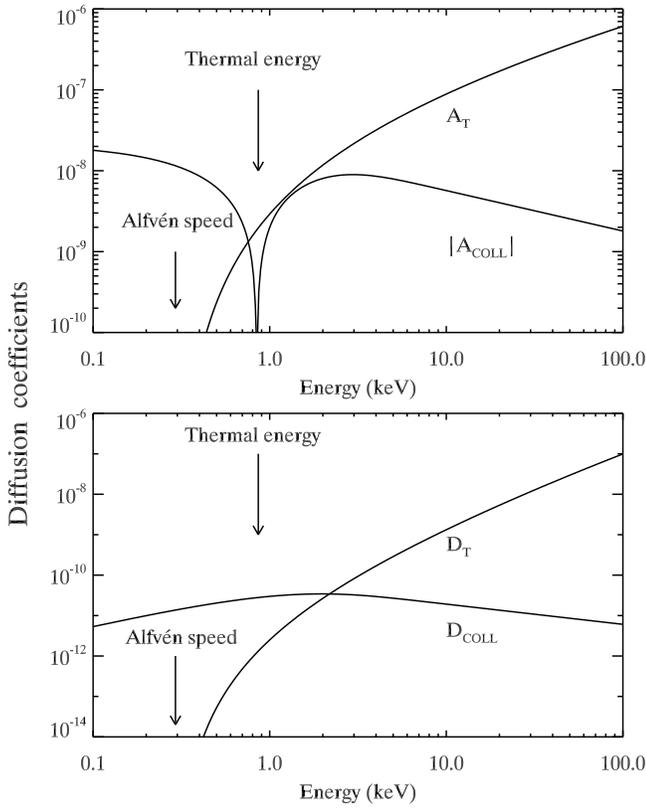}}
\caption{ 
  Upper panel: the absolute value of the dimensionless Coulomb (subscript
  $_\mathrm{COLL}$) and transit-time damping (subscript $_\mathrm{T}$)
  convection coefficients.  Lower panel: the dimensionless Coulomb and
  transit-time damping diffusion coefficients. The arrows mark the thermal energy
  of the electrons in the plasma and the kinetic energy of electrons moving with
  the Alfv\'en speed $v_A$.}
\label{fig:adcoeff}
\end{figure}

The coefficients for Coulomb collisions of the accelerated particles with the
thermal background population (Spitzer \cite{spitzer62}, Trubnikov
\cite{trubnikov65}) are given by :
\begin{eqnarray}
  A_\mathrm{COLL} & = & -\, \nu\,n\,
  \left[\psi(x)-\psi^{\prime}(x)\right]E^{-1/2}\\ 
  D_\mathrm{COLL} & = &  2\,\nu\,n\, \frac{\psi(x)}{x}
  E^{1/2} 
\end{eqnarray}
with 
\begin{eqnarray}
  & & x =  \frac{m_ec^2\,E}{k_\mathrm{B}\,T}\,,\qquad
  \nu =   \frac{\sqrt{8}\, \pi e^4 \,\log\Lambda\,T_\mathrm{H}}{ m_e^2 c^3}\\ 
  & & \psi(x) =  2\pi^{-1/2}\int_0^x\sqrt{t}\mathrm{e}^{-t}\,dt\,.
%
\end{eqnarray}
The energy dependence of the acceleration and collisional coefficients is shown
in Fig.~\ref{fig:adcoeff}.

%
%
%
%

The sink term is given by:
\begin{equation}
  \label{eq:escapeterm}
S(E)=\frac{T_\mathrm{H}\beta}{\tau},
\end{equation}
where $\tau$ is the escape time. This represents a purely kinematic escape for a
spatially homogeneous electron distribution (leaky-box escape). We can
understand this term in the following way: let's assume that the plasma is
contained in a cylinder with length $L$ and cross section $A$, having a
spatially homogeneous electron density $n$ with a total number of particles
$N_\mathrm{TOT}=\int n\,dV=n\cdot L\cdot A$ all having the same speed $v$.
After a time interval $dt$, the number of particles which have escaped the
cylinder (assuming free streaming without collisions) is $dN_\mathrm{TOT}=n\cdot
v\cdot A\cdot\,dt$.  The rate of change of the density is $dn/dt=n\cdot v / L$.
Defining the escape time $\tau=L/c$ yields Eq.~(\ref{eq:escapeterm}), where the
factor $T_\mathrm{H}=\Omega_\mathrm{H}^{-1}$ comes from the transformation of
the equation into dimensionless form.

In a more general case, where collisions have an effect, the particles cannot
stream freely out of the accelerator, and they will need a longer time to
escape. In this case, the escape time $\tau$ can be larger than $L/c$ and the
energy dependence on the collision frequencies will cause $\tau$ to vary with
the energy. Since we do not address here the physical details of the escape we
will assume for now that $\tau(E)$ is constant for all energies.

Under the influence of collisions, we are not allowed to use $L$ as the length
of the path traversed by a particle before it can leave the accelerator:
$L(E)=c\tau(E)$ denotes the effective length for particle escape. It is best to
think of the escape time $\tau$ as a general parameter describing the strength
of the particle trapping in the accelerator: the trapping becomes the more
efficient, the larger $\tau$.

The source term $Q(E)$ is needed to keep the number of particles in the
accelerator constant. Physically, these are electrons supplied by a return
current, needed to keep electrically neutral the region where the electrons are
accelerated. We assume that the electrons supplied by the return current
mechanism are ``cool'' in the sense that their distribution is comparable with
a thermal spectrum and with the temperature of the ambient plasma. Therefore we
have
\begin{equation}
  \label{eq:sourceterm}
  Q=\dot{n}_0\cdot N_\mathrm{MB}(E)\,,
\end{equation}
where
\begin{equation}
N_\mathrm{MB}(E)=2\pi^{-1/2}(k_\mathrm{B}T)^{-3/2}\sqrt{E}\mathrm{e}^{-E/(k_\mathrm{B}T)}
\end{equation}
is the Maxwell-Boltzmann distribution normalized to unity, and $\dot{n}_0=\int
SN\,dE$ is the rate of escaping particles.

%
%
\begin{figure}
\resizebox{\hsize}{!}{\includegraphics{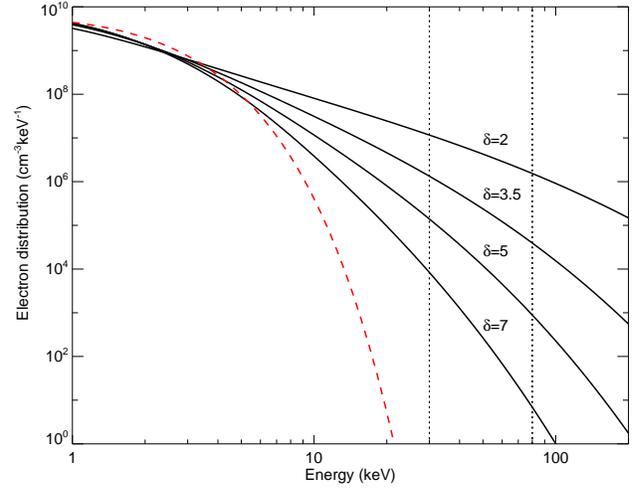}}
\caption{Accelerated electron distributions with different values of the
  power-law index resulting from changes in $I_\tau=I_\mathrm{ACC}\cdot\tau$.
  The dashed curve represents the ambient Maxwellian distribution. The two
  dotted lines indicate the energy range used for the computation of the
  power-law index $\delta$ shown above each spectrum.}
\label{fig:spectra_delta}
\end{figure}
%
%

\subsection{Method of Solution}
Equation (\ref{eq:maindiff}) is solved numerically using the Crank-Nicolson
finite differences scheme in logarithmic energy space, which is well suited for
diffusion problems. This method is the same as used in MLM and gives accurate
results with fixed steps in time and $\log\,E$.

\subsection{From Electrons to Photons}
The numerical solution described above yields electron spectra. For a meaningful
comparison with the observations we need photon spectra. They are generated by
bremsstrahlung from the electron distribution. Since the model spectra are in
equilibrium (particle losses by escape and coulomb collisions are compensated by
the acceleration) a \emph{thin-target} emission is computed.

We convert the energy differential electron flux distribution $F(\ele)$ in
electrons cm$^{-2}$ s$^{-1}$ keV$^{-1}$ into the photon spectrum $J(\phe)$
observed at Earth in photons cm$^{-2}$ s$^{-1}$ keV$^{-1}$ using the equation
for the thin-target bremsstrahlung emission:
\begin{equation}
J(\phe)\,\mathrm{d}\phe=\frac{nV}{4\pi\,R^2}
\int_\phe^\infty N(\ele)d\sigma(\phe,\ele)\,dE\,,
\end{equation}
where $V$ is the source volume and $R$ is the distance from the Sun. The cross
section $d\sigma(\phe,\ele)$ used is the fully relativistic, spatially integrated
Bethe-Heitler formula (Bethe \& Heitler \cite{BH1934}) without further
approximations. This should adequately represent the emission process of the
computed electron spectrum in a looptop source.

We will consistently use the notation $\delta$, $\gamma$, $F_{E_0}$,
$J_{\phe_0}$, $E_*$, $\phe_*$, $F_*$ and $J_*$ for, respectively, the electron
spectral index, the photon spectral index, the electron density at energy $E_0$,
the photon flux at energy $\phe_0$, the electron pivot-point energy, the photon
pivot-point energy, the electron density at $E_*$ and the photon flux at
$\phe_*$.

%

%
%
\begin{figure}
\resizebox{\hsize}{!}{\includegraphics{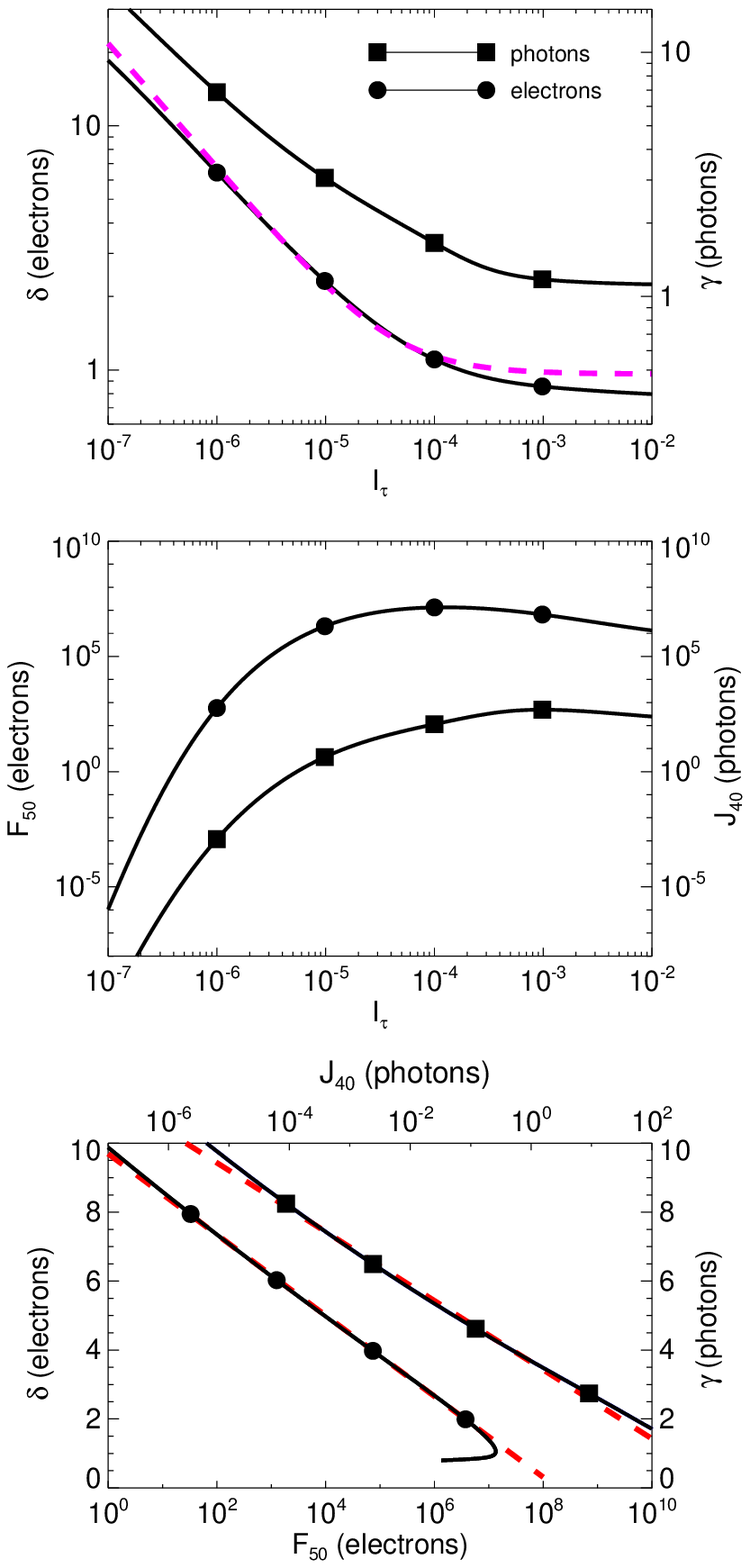}}
\caption{{\it Top:} The electron spectral index $\delta$ (measured in the energy
  range 30--80 keV) and the photon spectral index $\gamma$ (measured in the
  energy range 30--50 keV) plotted against $I_\tau$. The dashed line represent
  the approximated analytical solution computed in App. \ref{app:a}. \newline
  {\it Middle:} The electron density at 50 keV $F_{50}$ and photon flux at
  40~keV $J_{40}$ plotted against $I_\tau$. \newline
  {\it Bottom:} $\delta$ vs. $F_{50}$ and $\gamma$ vs. $J_{40}$ plotted
  parametrically as a function of $I_\tau$. The dashed line are the best log-lin
  fit for $\delta$ in the range 2--8 and for $\gamma$ in the range 3--9.}
\label{fig:deltaflux}
\end{figure}

%
%

\section{Results}
\label{results}

Section \ref{ss:resdef} presents the detailed properties of the numerical
solution of Eq.~(\ref{eq:maindiff}) for the standard set of parameters
(``default values'' given in Table \ref{tab:vardesc}) and the energy independent
escape model.  In a second step (Sect.  \ref{ss:parexp}) we will proceed to
study the dependence of the results on the values of the model parameter and,
finally, examine an alternative escape model (Sect. \ref{ss:altesc}).

\subsection{Results for the Default Values of the Model Parameters}
\label{ss:resdef}

In this section the default values for the model parameters presented in Table
\ref{tab:vardesc} are used. In this case, $\tau$ and $I_\mathrm{ACC}$ are the
only free parameters of the model, and the acceleration coefficients are much
larger than the Coulomb collisional coefficients in the energy range above 10
keV (see Fig.~\ref{fig:adcoeff}). Therefore, we expect that the electron
spectrum in this energy range depends only on the value of the product
$I_\tau=I_\mathrm{ACC}\cdot\tau$ (cf. Eq.~\ref{eq:maindiff}), and we can
basically work with one free parameter. For some choices of the model
parameters, this approximation will not be valid anymore, so we will relax this
assumption in Sect. \ref{ss:parexp}.

Figure~\ref{fig:spectra_delta} presents the electron spectra resulting from the
numerical solution of Eq.~(\ref{eq:maindiff}) for 4 different values of
$I_\tau$, which yield different values of the electron spectral index $\delta$
fitted in the energy range 30--80 keV. The harder spectra are the ones resulting
from a higher value of $I_\tau$. The dependence of the spectral index on
$I_\tau$ is shown in the top panel of Fig.~\ref{fig:deltaflux}. Note that in the
case where acceleration is weak or escape strong (that is, $I_\tau$ is small)
very soft spectra are produced. For increasing trapping and acceleration
efficiency ($I_\tau$ larger) the spectra become harder. The middle panel of
Fig.~\ref{fig:deltaflux} shows the dependency of the electron flux at 50 keV,
$F_{50}$, on $I_\tau$. Here the flux is larger for larger $I_\tau$. During a
flare, $I_\tau$ will change as more energy is injected into turbulence waves,
and therefore both $\delta$ and $F_{50}$ will change. This explains
qualitatively the soft-hard-soft effect: as $I_\tau$ increases, $\delta$ will
decrease and $F_{50}$ will increase. It is shown in the bottom panel of
Fig.~\ref{fig:deltaflux}, where $\delta$ vs. $F_{50}$ are plotted as a function
of the parameter $I_\tau$.

Since $I_\tau$ contains the physics of acceleration and escape, the time
evolution of the spectral index and flux in our model is a direct consequence of
the changes in $I_\tau$. For example, if the energy density of the turbulent
waves grows, reach a maximum value and then decreases, the electron spectrum
will harden until peak time and soften again.

To check the quality of the numerical solution, we obtain an approximated
analytical expression for the function $\delta(I_\tau)$, whose computation is
explained in Appendix \ref{app:a}. The approximate curve is plotted in the top
panel of Fig.~\ref{fig:deltaflux} as a dashed line and agrees well with the
numerical solution.  Because of the linear nature of Eq.~(\ref{eq:maindiff}),
the method exploited in App. \ref{app:a} cannot be used to get an approximate
solution for $F_{50}(I_\tau)$.

The dashed line in the bottom panel of Fig.~\ref{fig:deltaflux} represents the
best fit of a logarithmic function (linear in log-lin space) to the data for
$\delta$ in the range 2--8. Such a relation between the electron flux at a
reference energy and the spectral index indicates the presence of a pivot point
in the electron spectra,  as shown in section \ref{pivpointintro}. From the
slope and normalization of the dashed line we can get values of the pivot-point
energy $E_*$ and flux $F_*$ for the electron distributions.

More precisely, if
\begin{equation}
\delta=a\log F_{E_0}+b\,,
\end{equation}
then
\begin{equation}
 E_*=E_0\mathrm{e}^{1/a}\qquad\mbox{and}\qquad
 F_*=\mathrm{e}^{-b/a}\,
\end{equation}
as follows from Eq.~(\ref{abdef}). The dashed line in Fig.~\ref{fig:deltaflux},
bottom, corresponds to $E_*=7.65$~keV and $F_*=2\cdot 10^8$
electrons~cm~$^{-3}$~keV~$^{-1}$.

For a meaningful comparison of these results with the observations, a value for
the pivot point of the \emph{photon spectra} is needed. $J_*$ and $\phe_*$ can
be computed by fitting the regression line of $\gamma$ vs. $\log J_{\phe_0}$, in
the same way as $E_*$ and $F_*$. Note that while $J_*$ depends on the source
volume, $\phe_*$ can be immediately compared with the observations. Using the
photon spectral index computed in the energy range 30--50~keV, the regression
(restricting $\gamma$ to the range 3--9) yields the pivot-point coordinates
$\phe_*=5.0$~keV and $J_*=1.9\cdot 10^3\cdot V_{27}$ photons cm$^{-2}$ s$^{-1}$
keV$^{-1}$, where $V_{27}$ is the volume of the source in units of
$10^{27}$~cm$^3$.

%
%
\begin{figure}
\resizebox{\hsize}{!}{\includegraphics{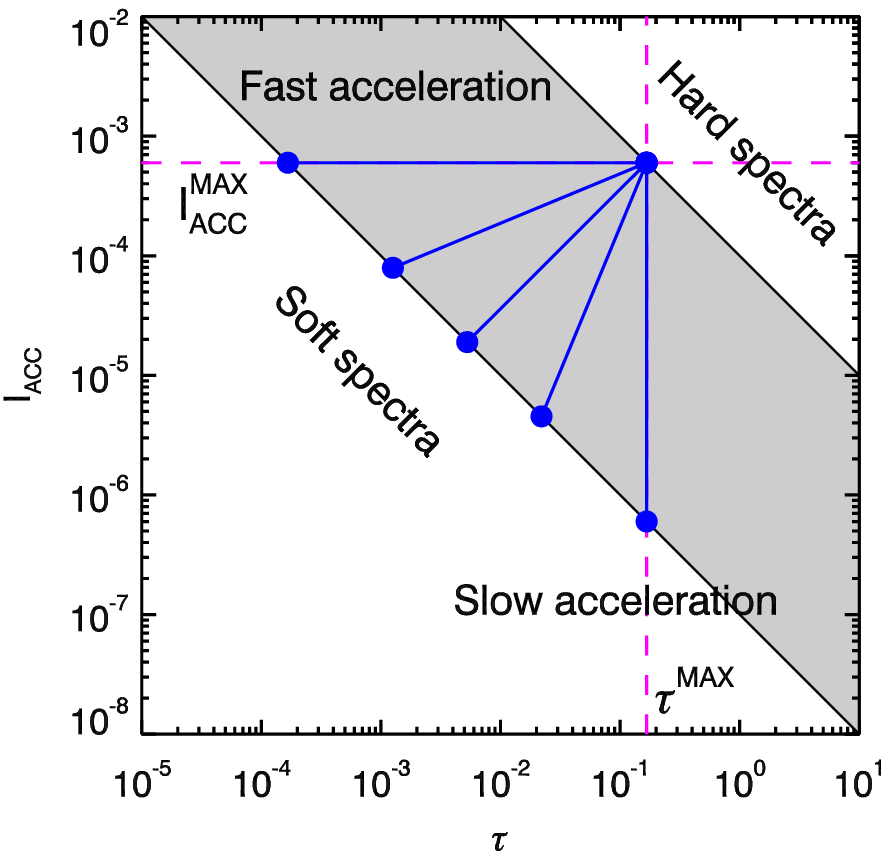}}
\caption{Five paths (between filled circles) in the $\tau,I_\mathrm{ACC}$--plane
  as described by Eq.~(\ref{eq:pathalpha}), corresponding to
  $\arctan\alpha=0,\pi/4,\pi/2,3\pi/4,\pi/2$. The gray shaded region contains
  the points yielding photon spectra with a spectral index lying approximatively
  in the 1.5--15 range. The two dashed lines have abscissa and ordinate of
  $\tau^\mathrm{MAX}$ and $I_\mathrm{\,ACC}^\mathrm{\,MAX}$, respectively.}
\label{fig:plot}
\end{figure}

%
%

The above value of the pivot-point energy is too low compared to the looptop
source observations by Battaglia \& Benz (\cite{battaglia06}) summarized in
Table \ref{tab:obsres}, reporting $\phe_*\approx 20$~keV. This means that if the
photon spectral index of our model varies from, say, 3 to 6, the non-thermal
flux is subjected to an excursion which is too large to account for the observed
behavior of the spectrum in solar flares.

Two possible solutions of this problem exist: we can explore the parameter space
to find out if there is some combination of the parameter values which indeed
produces a higher pivot-point energy $\eps_*$, or we can try to modify the
escape term.

\subsection{Exploration of the Parameter Space}
\label{ss:parexp}

The set of default parameters used in the previous section yields a pivot-point
energy $\phe_*$ which is lower than the observed value.  However, the
pivot-point energy depends on model parameters like $T$, $n$ etc.  Therefore a
better coverage of parameter space than the example reported above is needed to
assess the range of variability of $\phe_*$. The pivot-point energy depends on
the temperature of the ambient plasma $T$, its electron density $n$ and its
magnetic field strength $B_0$ (indirectly, through the Alfv\'en speed).  We
therefore compute the equilibrium solutions of Eq.~(\ref{eq:maindiff}) for a
large number of combinations of different values of these parameters, each set
yielding a value for $\phe_*$.  In this way we sample the function
$\phe_*(T,n,\cdots)$.

%
%
%
\begin{figure}
\centering
\resizebox{\hsize}{!}{\includegraphics{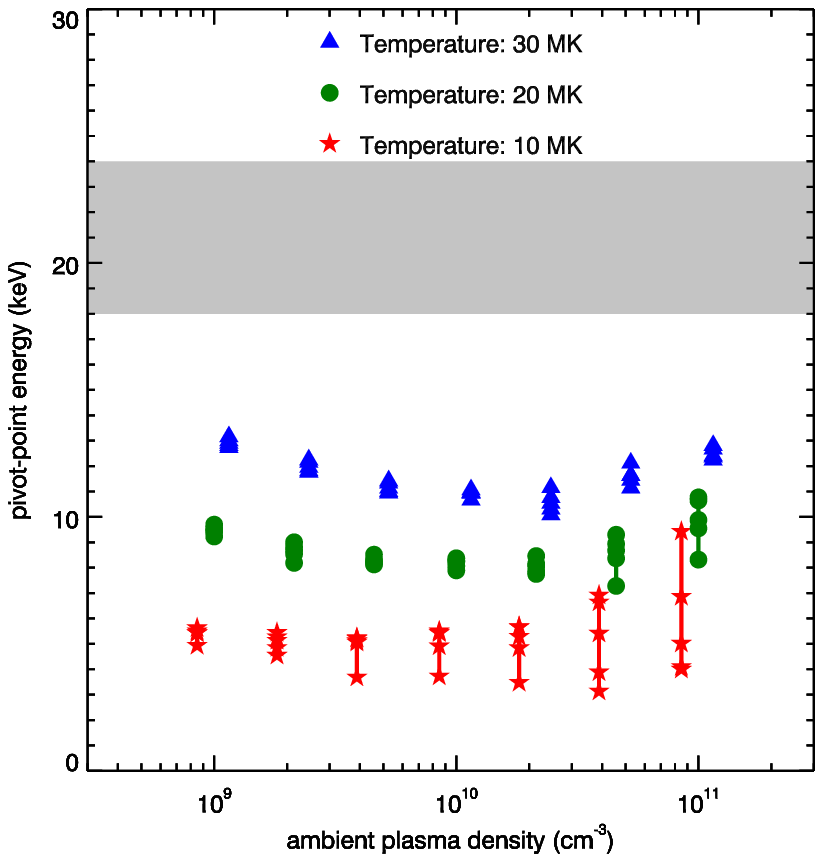}}
\caption{Dependance of the photon pivot-point energy $\eps_*$ on the electron
  density $n$ for plasma temperatures $T$ of 10 MK (stars), 20 MK (circles) and
  30 MK (triangles). The density values of the 10 and 30 MK data points have
  been shifted by 15\% to avoid overlap with the other points. The spread arise
  from computation of the spectra along different paths as explained in the
  text. The gray shaded region represents the observed range of values of
  $\eps_*$ reported in Table~\ref{tab:obsres}.}
\label{fig:explore1}
\end{figure}

%
%

There is however a complication: for some choices of the parameters, the
approximation that the spectral index and flux depend only on $I_\tau=\tau\cdot
I_\mathrm{ACC}$ will not be valid anymore. This approximation becomes invalid if
$A_\mathrm{COLL}/A_\mathrm{T}$ or $D_\mathrm{COLL}/D_\mathrm{T}$ approach unity.
We note that both ratios are roughly proportional to~$n^2$ at high energies, and
therefore the approximation becomes bad for large values of the density.  In
this case, the escape time $\tau$ and the acceleration parameter
$I_\mathrm{ACC}$ separately influence the spectral index and flux, and the
pivot-point energy will depend on the $\tau$ vs. $I_\mathrm{ACC}$ relation.
Since we have not modeled the physical mechanism responsible for the trapping of
particles, the details of how $\tau$ depends on $I_\mathrm{ACC}$ are not known.

We estimate the range of pivot points by choosing several representative paths
in the $\tau,I_\mathrm{ACC}$--plane (as shown in Fig.~\ref{fig:plot}) and
computing the pivot-point coordinates resulting from spectra computed along
these paths.

Recalling that in first approximation the spectral index depends only on the
product $I_\tau=I_\mathrm{ACC}\cdot \tau$, the paths are required to start from
a point on the line $\tau\cdot I_\mathrm{ACC}=10^{-7}$ and end in a point on the
line $\tau\cdot I_\mathrm{ACC}=10^{-4}$. The region between these two
delimitation lines is shown in gray in Fig.~\ref{fig:plot}. These limits are
chosen to ensure that the computed photon spectral indices fall into the
observed range. For the standard values of the parameters, the region above the
gray area yields spectral indices harder than about 1.5 and the region below
yields $\gamma$ softer than about 15.

There are however more constraints on the values of $I_\mathrm{ACC}$. This
parameter is the product of the dimensionless wave number and the energy density
in the turbulent fast mode waves. In our simplified version of transit-time
damping acceleration, the time evolution of the spectral energy distribution of
the waves is not computed. Therefore we set the maximum value of
$I_\mathrm{ACC}$ at $I_\mathrm{\,ACC}^\mathrm{\,MAX}=6\cdot 10^{-4}$. It is the
maximum value that is obtained by MLM during the cascading process (read from
the plot in Fig.~6b in MLM). Since $I_\mathrm{ACC}$ controls the acceleration
efficiency, values which are much smaller then $I_\mathrm{ACC}^\mathrm{\,MAX}$
yield a weak and slow acceleration, contrary to the observations. Larger values
of $I_\mathrm{ACC}$ could result from a higher energy density in the waves, but
in that regime the wave amplitude becomes so large that nonlinear effects are
likely to a lead to a breakdown of the physical model used by MLM to compute the
acceleration coefficients. Thus values of $I_\mathrm{ACC}$ much larger than
$I_\mathrm{\,ACC}^\mathrm{\,MAX}$ do not necessarily improve the acceleration
efficiency.

%
%
%
\begin{figure*}
\centering
\resizebox{\hsize}{!}{\includegraphics{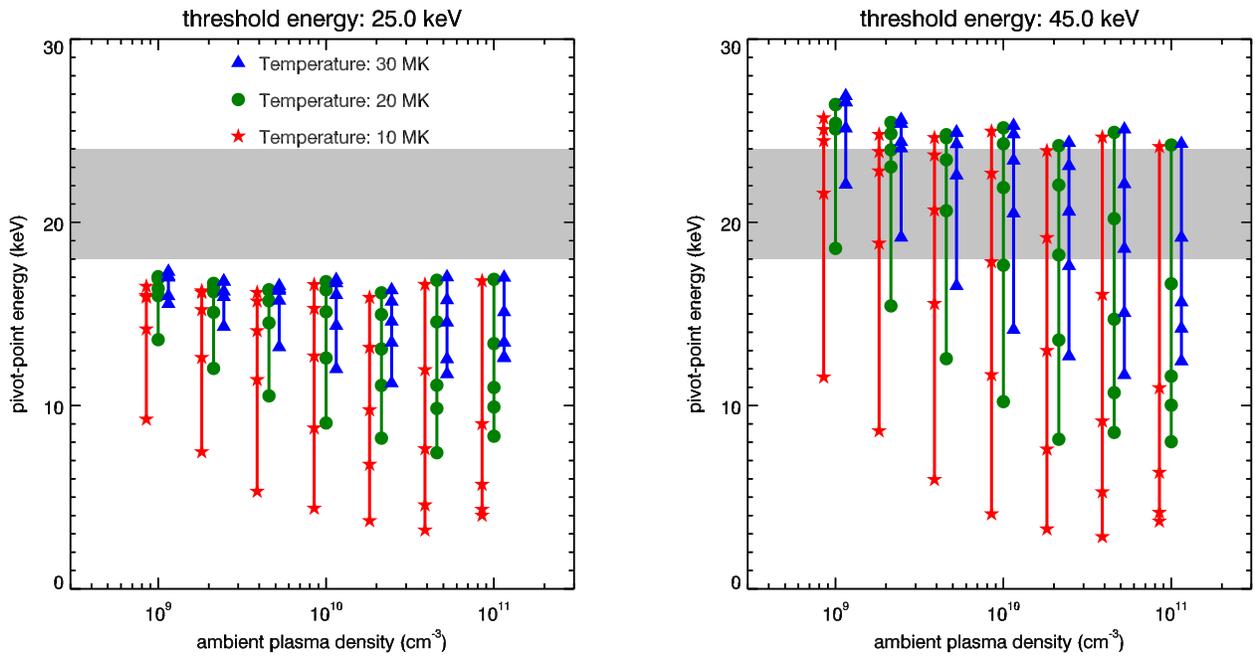}}
\caption{Same as in Fig.~\ref{fig:explore1}, but with a modified escape
  term featuring perfect trapping below 25 keV (left panel) and below 40 keV
  (right panel).}
\label{fig:explore2}
\end{figure*}

%
%

For these reasons we additionally require that all the paths end on the point
$(\tau^\mathrm{\,MAX},I_\mathrm{\,ACC}^\mathrm{\,MAX})$, where
$\tau^\mathrm{\,MAX}$ is the point on the upper delimitation line with ordinate
$I_\mathrm{\,ACC}^\mathrm{\,MAX}$. Furthermore, we also require the paths to be
monotonically increasing functions of both coordinates (this guarantees that the
product $I_\mathrm{\tau}$ is also monotonically increasing). Our reference paths
are thus given by
\begin{equation}
  \label{eq:pathalpha}
  I_\mathrm{ACC}=I_\mathrm{ACC}^\mathrm{\,MAX}
  \left(\frac{\tau}{\tau^\mathrm{\,MAX}}\right)^{\alpha}
\end{equation}
for different values of $\alpha$ between 0 and $\infty$. They represent straight
lines in the logarithmic $I_\mathrm{ACC}$ vs. $\tau$ plot shown in
Fig.~\ref{fig:plot}, intersecting in the point with coordinates
$(\tau^\mathrm{\,MAX},I^\mathrm{\,MAX}_\mathrm{ACC})$.

The results are shown in Fig.~\ref{fig:explore1}, where the value of the
pivot-point energy $\eps_*$ is plotted for 3 different values of the plasma
temperature ($T=10,20,30$~MK) and several different values of the plasma density
between $n=10^9\,\mathrm{cm}^{-3}$ and $n=10^{11}\,\mathrm{cm}^{-3}$. The
different points for each temperature and density represent the range of
variation of~$\eps_*$ for different values of $\alpha$ as explained above. We
have used $\alpha=0,\sqrt 2-1,1,\sqrt 2+1,\infty$.

The pivot-point energy increases with temperature. The density dependence of
the pivot-point energy is weak, while the pivot-point flux decreases at higher
densities. The number of accelerated particles becomes very small at densities
of $10^{12}$~cm$^{-3}$ and higher, because the particles lose energy in the
collisions faster than they gain energy from the waves. The spread of the
pivot-point energies shown in Fig. \ref{fig:explore1} increases at higher
densities, as expected from the high-density breakdown of the approximation that
the pivot point only depends on $I_\mathrm{ACC}\cdot \tau$.

\subsection{Alternative Escape Modeling}
\label{ss:altesc}

As an alternative way for explaining the high values of $\phe_*$, we now modify
the escape mechanism. For instance, if an electric potential $V_\mathrm{E}$ is
present between the accelerator and the footpoints, electrons with kinetic
energy lower than the threshold energy $E_\mathrm{T}=eV_\mathrm{E}$ will not be
able to cross that barrier and will therefore not leave the accelerator. Such an
electric field is expected to exist and drive the return current of electrons
from the chromosphere. In this scenario the trapping is ideal below
$E_\mathrm{T}$, and $\tau$ becomes infinitely large.

This means that the escape time is given by:
\begin{equation}
\tau(E)=\left\{ \infty\quad\mbox{if}\quad E\le E_\mathrm{T}  \atop 
                \tau  \quad\mbox{if}\quad E>   E_\mathrm{T} 
        \right.
\end{equation}

The results are shown in Fig.~\ref{fig:explore2} for two values
of~$E_\mathrm{T}$, 25~keV and 45~keV. Note that the value of the pivot-point
energy $\phe_*$ increases with increasing $E_\mathrm{T}$. The pivot point
energies reach the observed values for $E_\mathrm{T}\simeq 30$~keV, although
even at 45 keV some paths in the $\tau,I_\mathrm{ACC}$--plane deliver lower
pivot-point energies than observed.

The points in Figs.~\ref{fig:explore1} and~\ref{fig:explore2} with larger
pivot-point energy correspond to the paths with the lowest values of $\alpha$,
where the escape time $\tau$ changes faster than the acceleration parameter
$I_\mathrm{ACC}$.

\begin{figure*}
\centering
\resizebox{\hsize}{!}{\includegraphics{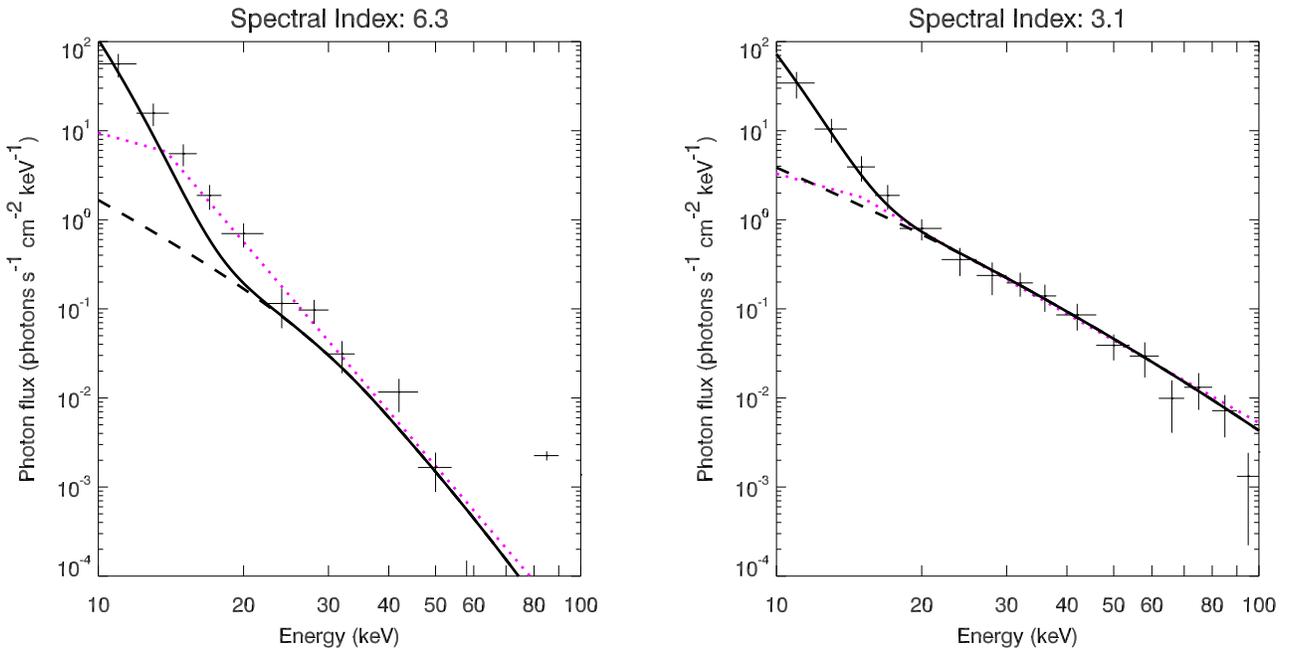}}
\caption{Looptop source spectra observed during the M3 flare of December 4,
  2002, represented by crosses (bin width and error bar, data courtesy of M.
  Battaglia). The dotted line is a power-law fit to the data (a low-energy
  turnover at around 15 keV) with spectral index $\gamma$ of 6.3 for the {\it
    left} panel and 3.1 for the {\it right} panel. In each panel, the dashed
  line is a stochastic acceleration model spectrum computed for model
  parameters: temperature $T=20$~MK, density
  $n=5\cdot10^{10}\,\mathrm{cm}^{-3}$, and threshold energy
  $E_\mathrm{T}=40$~keV (chosen such that the pivot-point energy $\phe_*$
  matches the observed value of $\sim$~18~keV reported in Battaglia \& Benz
  \cite{battaglia06}). The continuous line is the sum of the model spectrum with
  an additional isothermal component, needed to fit the low-energy range of the
  spectra.}
\label{fig:overlay}
\end{figure*}

A comparison of observed spectra of a looptop source (for the M3 flare of
December 4, 2002) and the spectra computed by the stochastic acceleration model
with $E_\mathrm{T}=40$~keV is shown in Fig. \ref{fig:overlay}. The model
parameters used were temperature $T=20$~MK and density
$n=5\cdot10^{10}\,\mathrm{cm}^{-3}$ and $\alpha=\sqrt{2}-1$, yielding a
pivot-point energy $\eps_*=18.7$~keV, similar to observed value of $18.1$~keV
reported by Battaglia \& Benz (\cite{battaglia06}). The model spectra are
represented by the dashed lines, the power-law fit to the data by the dotted
line, and the sum of the model spectra with an isothermal component by the
continuous line.

In our model, the value of the total photon flux observed at earth depends
linearly on the accelerator volume, which acts as a normalization factor for the
model spectra. For the comparison with the observed data, the volume can be
freely chosen, but is assumed not to change along the path in the
$\tau,I_\mathrm{ACC}$--plane. In practice, that means that the volume can be
chosen in order to match the data (or the power-law fittings) at, say, peak
time, and automatically the spectra at all the times during the emission peak
will match, provided that the pivot-point energies of the observed and the model
spectra are the same.

The low energy part of the observed spectrum shows a thermal emission much
larger than the model spectra emission. This can be understood if particle
acceleration takes place in a smaller volume than the one where the thermal
emission takes place. For the events shown in Fig. \ref{fig:overlay}, the
filling factor of the accelerator amounts to around $10^{-3}$.

The left panel of Fig. \ref{fig:overlay} shows a mismatch between the observed
data and the model spectra below 30~keV. This is due to the effect of the escape
model used: the suppression of electron escape results in a hardening of the
electron spectra below $E_\mathrm{T}=40$~keV. Particles with energy lower than
$E_\mathrm{T}$ are able to extract more energy out of the waves until they are
accelerated beyond $E_\mathrm{T}$. A turnover occurs at around $0.7
E_\mathrm{T}$ for photon spectra, due to the shape of the bremsstrahlung cross
section $d\sigma(\phe,\ele)$.

%
%
%

%
%

%
\section{Discussion}
\label{discussion}

The important results are the following:
\begin{itemize}
\item The transit-time damping mechanism with escape is able to provide photon
  spectra with spectral indices in the deka-keV energy range corresponding to
  the observed range for flares ($\gamma=$ 2 to 10, Dennis \cite{dennis85},
  Battaglia et al.  \cite{battaglia05}) as the ratio between the trapping and
  accelerating efficiency is varied (Fig. \ref{fig:spectra_delta}).  This
  confirms that escape indeed does soften the spectrum for different values of
  densities and temperatures of the accelerator.
\item Assuming constant values of temperature and density in the accelerator (as
  may be the case during a hard X-ray peak with a duration of some tens of
  seconds) we are able to produce a correlation between the photon spectral
  index and flux yielding an approximate pivot point in photon flux vs.  energy
  space (Figs.~\ref{fig:deltaflux} and \ref{fig:explore1})
\item The energy of the pivot point in the simplest model is around 10 keV and
  is lower than the reported observed value of 20 keV.
\item An additional increase of the strength of trapping below 20-30 keV (as may
  be obtained from the presence of a potential barrier) increases the
  pivot-point energy to 20 keV (Fig. \ref{fig:explore2}).
\item The potential barrier hardens the spectrum below its threshold energy.
  For large $\gamma$, this flattens the spectrum more than observed in the range
  where the non-thermal emission still dominates the thermal emission
  (Fig.~\ref{fig:overlay}, left).
\item The observational upper limit for the energy of a turnover in the photon
  spectra lies around 15~keV (Saint-Hilaire \& Benz \cite{saint-hilaire05}). 
  Model spectra computed for threshold energies below about 23~keV yield
  a turnover lower than 15~keV and pivot-point energies below 16~keV.
\end{itemize}
  What is the physical significance of the pivot point? We would like to point
  out first that there is not necessarily any: it can be thought of as a useful
  parameter which describes the variation in flux and spectral index of the
  observed spectra, and one has to explain this variation first in the context
  of a model such as done in this work. However, the fact that it appears at an
  energy which is comparable to the estimated turnover energy of the non-thermal
  component, as well as the energy below which the thermal component dominates
  is a suggestive fact.

  The model suggests that the main factor influencing the pivot-point energy is
  the number of electrons available at energy comparable with it. We propose
  that, in general, the presence of a large supply of electrons at some energy
  implies a similar energy for the pivot point, on the condition that the
  electron collisional energy losses are negligible above that energy range,
  such that the acceleration process can effectively energize that population.
  It is not important if such a population exist at the beginning of the
  acceleration, but it must be present in the later stages and be maintained
  until the equilibrium is reached.

%

\section{Conclusions}
\label{conclusion}

The model studied assumes that turbulent fast-mode waves are present in the
accelerator region. The equilibrium spectrum of electrons subjected to Coulomb
collisions and transit-time damping interactions with the waves are computed
allowing for trapping/escape and replenishment of ``cold'' particles. As such
the model does not distinguish between the accelerator and the radiator and
considers the photon spectra emitted by the electrons in the accelerator itself.
The relatively dense looptop sources observed by RHESSI fit this scenario as
accelerators/emitters, and the observed high densities support a scenario in
which trapping plays an important role in acceleration.  Therefore, spectral
observations of these looptop sources may well deliver a snapshot of the
electrons in the accelerator.

To match the low-energy part of the observed spectrum, we have to add an
additional thermal component at the observed temperature. This component fills a
much larger volume (factor of 1000). The heating may have occurred by earlier
acceleration or may be the result of waves escaping from the accelerator region
with energy density too low to yield significant energy gain per particle.

The goals of this work were to test whether a stochastic acceleration mechanism
can account for the observed soft-hard-soft behavior and, in particular, to
find a minimal set of modifications to the transit-time damping mechanism
yielding a pivot-point like behavior of the electron spectrum in the deka-keV
range.  Therefore we have not fully investigated the physics of escape, and in
particular we leave open the question of the connection between the acceleration
process and the trapping mechanism.

The results shown in Section \ref{results} confirm that the hardness of the
spectrum of accelerated particles in a stochastic acceleration model depends
strongly on particle trapping. This can be understood by recalling that
stochastic acceleration can be thought as a transport of particles in energy
space due to a random walk process. While acceleration pushes particles toward
higher energies, the escape acts against this flow, because fast particles are
more likely to be lost, and replaced by particles at lower energies. Therefore,
when the losses are stronger, the average time a particle spends random-walking
in energy space is shorter, thus gaining less energy. This results in softer
spectra.

We have used one specific kind of stochastic acceleration model and a simple
escape scenario, but the physical mechanisms explored by the specific models are
general enough that other stochastic models and escape terms are expected to
follow the same general behavior leading to a soft-hard-soft effect, manifesting
itself as a correlation of the time evolution of the spectral index and the
non-thermal X-ray flux.

A simple escape model cannot account for the observed value of the photon
pivot-point energy, which is about a factor 2 higher. The modification in the
escape term featuring no escape below a threshold energy $E_\mathrm{T}\simeq
40$~keV can increase the energy of the pivot point to the observed values, but
at the price of introducing a spectral hardening below 30~keV, which is not
observed for large $\gamma$.

One possible interpretation of the threshold energy used in the alternative
escape model is the presence of an electric field driving the return current.
It was treated as a free parameter, but physically the escape process and the
return current are linked by the electric conductivity of the loop legs.  In
this sense the model is not self consistent. The issues of transport of the
escaped particles to the footpoint and return currents are important (see for
example Zharkova \& Gordovsky \cite{zharkova05}) and need to be addressed in
future work to be able to derive footpoint spectra.

Thus the model is not able to account quantitatively for all the observed
features of the intricate spectral behavior of the hard X-ray emission from
looptop sources of solar flares. Nevertheless, the model qualitatively accounts
for the key features of the spectral evolution, going one step further toward
the solution of the acceleration problem. This allows to interpret the
impressive RHESSI observations showing spectral variation in looptop sources.

The simplifications introduced by the model may well be responsible for its
shortcoming in reproducing the observed spectra for large $\gamma$. Future
improvements may include an isotropization process and magnetic trapping. This
will require the extension of the model including at least one spatial dimension
and the pitch angle distribution of the particles.


\begin{acknowledgements}
  We thank M. Battaglia for sharing her results before publication, and K.
  Arzner, S. Bruderer, L. Vlahos and the participants of the theoretical
  working group at the 6th RHESSI Workshop for useful discussions. The analysis
  of RHESSI data at ETH Zurich is partially supported by the Swiss National
  Science Foundation (grant nr. 20-67995.02). This research has made use of
  NASA's Astrophysics Data System Bibliographic Services.
\end{acknowledgements}

\appendix
\section{Approximate analytical solution}\label{app:a}
We compute here an approximate analytical solution in the energy range
$kT/(mc^2)\ll E \ll 1$. In this range, we can neglect the influence of the
Coulomb collisional coefficients $A_\mathrm{C}$ and $D_\mathrm{C}$ as well as
the source term $Q$. In equilibrium Eq.~(\ref{eq:maindiff}) becomes:
\begin{equation}
\label{eq:diffsimp}
\frac{D_0}{2}\difff{N}{E}+\left(D_1-A_0\right)\diff{N}{E}+
\left(D_2-A_1-S\right)N=0\,,
\end{equation}
where the coefficients $A_i$ and $D_i$ are the factors of order $i$ occurring in
the Taylor expansion of $A$ and $D$ around $E$. Their computation is tedious,
but straightforward. Neglecting the term of order $\xi^2$ and setting
$\gamma=1$ we get
\begin{eqnarray}
D_0 & = & K\beta^3\left(-{5}/{4}-\log\xi\right)\\
D_1 & = & K\beta \left(-{11}/{4}-3\log\xi\right)\\
D_2 & = & \frac{2K}{\beta}\left(-{17}/{8}-{3}/{2}\cdot\log\xi\right)\\
A_0 & = & 2K\beta\left(-1-\log\xi\right)\\
A_1 & = & \frac{2K}{\beta}\left(-{1}/{4}-\log\xi\right)
\end{eqnarray}
where $K=\frac{\pi}{8}\beta_\mathrm{A}^2\,I_\mathrm{ACC}$,
$\xi=\beta_\mathrm{A}/\beta$.

The solution $N(E)$ will be approximatively given by a
power-law around $E=E_0$
\begin{equation}
N(E)=N_0\left(\frac{E}{E_0}\right)^{-\delta}\,.
\end{equation}
Plugging this function into Eq.~(\ref{eq:diffsimp}) yields a quadratic equation
for $\delta$:
\begin{equation}
c_2\delta^2+c_1\delta+c_0=0
\end{equation}
with coefficients
\begin{eqnarray}
c_2 & = & \frac{\tau D_0}{2E_0^2} \\
c_1 & = & \frac{\tau}{E_0}\left(\frac{D_0}{2E_0}-D_1+A_0\right) \\
c_0 & = & \tau D_2-\tau A_1-\sqrt{2E_0}
\end{eqnarray}

The solution of the quadratic equation yields $\delta$ as a function of
$I_\tau=I_\mathrm{ACC}\cdot\tau$ and the model parameters.
The asymptotic behavior of the solution is the following:
$\delta$ is constant for $I_\tau\rightarrow\infty$, and 
$\delta$ is proportional to $1/\sqrt{I_\tau}$
for $I_\tau\rightarrow 0$.

%

%

\begin{thebibliography}{}
%

\bibitem[2002]{aschwanden02}
Aschwanden, M.~J. 2002, Space Science Reviews, 101, 1 

\bibitem[2006]{battaglia06}
Battaglia, M \& Benz, A.~O. 2006, A\&A, in press 

\bibitem[2005]{battaglia05}
Battaglia, M., Grigis, P.~C., \& Benz, A.~O. 2005, \aap, 439, 737 

\bibitem[2002]{benz02}
Benz, A.~O., \& Grigis, P.~C. 2002, \solphys, 210, 431 

\bibitem[1934]{BH1934} Bethe, H., \& Heitler, W. 1934,
Royal Society of London Proceedings Series A, 146, 83 
 
\bibitem[1996]{brosius96} Brosius, J.~W., Davila, J.~M., 
Thomas, R.~J., \& Monsignori-Fossi, B.~C. 1996, \apjs, 106, 143 

\bibitem[1985]{dennis85}
Dennis, B.~R. 1985, \solphys, 100, 465 

\bibitem[2003]{emslie03} 
Emslie, A.~G., Kontar, E.~P., Krucker, S., \& Lin, R.~P. 2003, 
\apjl, 595, L107 
 
\bibitem[2004]{emslie04}
Emslie, A.~G., et al. 2004, Journal of Geophysical Research
(Space Physics), 109, 10104 

\bibitem[2004]{grigis04}
Grigis, P.~C. \& Benz, A.~O. 2004, A\&A, 426, 1093

\bibitem[2005]{grigis05}
Grigis, P.~C., \& Benz, A.~O.\ 2005, \aap, 434, 1173 

\bibitem[2003]{holman03}
Holman, G.~D., Sui, L., Schwartz, R.~A. \& Emslie, G.~A. 2003, ApJ, 595, L97


\bibitem[2005]{kane05}
Kane, S.~R., McTiernan, J.~M., \& Hurley, K. 2005, 
\aap, 433, 1133 

\bibitem[1993]{1993}
Larosa, T.~N., \& Moore, R.~L. 1993, \apj, 418, 912 
 
\bibitem[2002]{lin02}
Lin, R.~P., Dennis, B.~R., Hurford, G.~J., et al. 2002, Sol. Phys., 210, 3

\bibitem[2003]{lin03}
Lin, R.~P., et al. 2003, \apjl, 595, L69 


\bibitem[2004]{liu04}
Liu, W.,  Jiang, Y.~W., Liu, S., \& Petrosian, V. 2004, \apjl, 611, L53 

\bibitem[1994]{Masuda94}
Masuda, S., Kosugi, T., Hara, H., et al. 1994, \nat, 371, 495
 
\bibitem[1996]{miller96}
Miller, J.~A., Larosa, T.~N., \& Moore, R.~L. 1996, ApJ, 461, 445 

\bibitem[1998]{miller98}
Miller, J. A. 1998, \ssr, 86, 79

\bibitem[2004]{petrosian04}
Petrosian, V.~\& Liu, S. 2004, ApJ, 610, 550 

\bibitem[2005]{saint-hilaire05}
Saint-Hilaire, P., \& Benz, A.~O.\  2005, \aap, 435, 743

\bibitem[2005]{sui05} 
Sui, L., Holman, G.~D., \& Dennis, B.~R.\ , 2005, \apj, 626, 1102

\bibitem[1962]{spitzer62} Spitzer, L.\ Physics of Fully 
Ionized Gases, New York: Interscience Publishers, 1962  

\bibitem[1965]{trubnikov65}
Trubnikov, B.~A., 1965, Reviews of Plasma Physics, 1, 105

\bibitem[2005]{zharkova05}
Zharkova, V.~V., \& Gordovsky, M. 2005, ESA SP-600: ``The 
Dynamic Sun: Challenges for Theory and Observations'', 147.1


\end{thebibliography}
\end{document}